# SEE-THROUGH REFLECTIVE OPTICAL ELEMENTS WITH EMBEDDED WAVELIKE FILMS

A.M. Smolovich[1], V.V. Kashin[1], V. Chernov[2]

[1]Kotel'nikov Institute of Radioengineering and Electronics, Russian Academy of Sciences, Moscow, Russia, 125009

[2]Universidad de Sonora, Hermosillo, Sonora, México, 83000

**Abstract:** The recently proposed optical elements containing semitransparent wavelike films embedded into the transparent material are investigated. Such optical elements do not distort a wave transmitted through them. Novel optical elements can be fabricated using well-known embossing methods. Their possible applications are discussed. The optical elements containing holographic wavelike films can be used for color holography, for augmented reality wearable glasses, as a beam combiner for head-up displays. The optical elements containing diffuse wavelike films can be used as an alternative for tinted car windows, instead of mirror-glass windows in high-rise buildings, for shop-windows, for one-way privacy windows, for printing images on eyeglasses. The scheme of a waveguide display with a single-layer holographic wavelike films is analyzed. The feasibility of the proposed applications is verified by estimation of the optical elements' parameters and computer simulation, and measurements conducted on an experimental specimen.

## 1. Introduction

Traditional reflection holograms and diffraction optical elements (OEs) with surface phase relief usually contain a thin metal layer deposited onto their relief for increasing the diffraction efficiency. It is possible to make these OEs semitransparent by choosing an appropriate thickness for the metal layer. However, a wave transmitted through these semitransparent OEs will acquire a phase modulation and, consequently, its wavefront will be distorted. There are several practical applications where a substantial reduction of this distortion is necessary. There are, for instance, head-up and head mounted displays (HUD and HMD, respectively). The reduction of this distortion can be done by using novel OEs containing semitransparent wavelike films (WFs) embedded into the transparent material that were analyzed in our previous paper [1]. The WF has a constant thickness and is bounded by surfaces with the same spatial relief. If a plane light beam is directed on the OE, then a part of the beam is reflected by a semi-transparent WF while the other part passes through it. The optical path lengths of the rays passing through different domains of the OE will be almost equal, since the refractive indices of the OE transparent material on both sides of the WF are the same. Therefore, the phase modulation of the wave transmitted through the OE should be small. This phase modulation will be estimated in our paper for the different types of WFs. Two types of the OEs will be considered, those that are holographic (locally periodic) and those that have a chaotic diffuse relief shape. Both types can include single layer as well as multilayer WFs.

The basic idea of an OE containing semitransparent WFs that do not distort the transmitted wave is simple. Previously, it was used in several patents (see, for example, patents [2,3]). The idea was especially useful for read-only optical discs containing several semitransparent information layers [4,5]. When information is read from one of the layers, the other layers do not distort the read or signal beams. A special example of an OE containing a semitransparent thin film that does not distort transmitted light is an interference structure recorded in a bulk medium using ultrashort laser pulses propagating towards each other. Such structures demonstrate a geometrical-optical mechanism of wavefront reconstruction, which is different from the holographic one [6,7].

However, as far as we know, the distortion of light passing through an OE containing holographic and chaotic diffuse WFs was not especially investigated before our publication [1]. It was obvious that it was necessary to apply some restrictions on the parameters of the WF, including its thickness, in order to ensure an acceptably low level of phase modulation of the transmitted light. It was important to understand what values of the diffraction efficiency of the wavelike hologram and its spectral selectivity can be achieved under these constraints. In [1], we evaluated the phase modulation of light transmitted through OEs containing various types of WFs and developed a dynamic theory of light diffraction for a multilayer wavelike hologram.

Novel OEs can be fabricated using well-known methods. For example, for the fabrication of the OE containing a holographic single-layer WF the following procedure should be used. Firstly, a hologram is recorded on photoresist. After photoresist development a common hologram with phase relief is obtained. This is a master hologram. Then, some metal should be deposited onto its relief surface. Then, the master stamp is produced via electrotyping. The master stamp can be made from nickel, nickel-cobalt, etc. This master stamp is used to emboss the surface relief onto the polymer layer. Then, a thin metal coating is deposited upon the embossed relief. Then, the same polymer material is placed on top. We see that these operations are usually used for mass-production of embossed holograms [8],[9, pp. 183-188].

As the mass production of the proposed OEs should not entail any significant difficulties, their applications are interesting. In this paper, the possible applications of OEs containing semitransparent WFs are discussed. The feasibility of the applications is verified with the appropriate estimations, computer simulations, and measurements executed on an experimental specimens. In the last section of the article, we discuss which types of OEs containing WF are the most promising in terms of their application, and compare them with the OEs that are currently in use. Special attention is paid to the possibility of using new OEs in waveguide-type displays.

## 2. OE with holographic single-layer WF

The OE shown in Fig. 1a represents a plane-parallel slab of transparent material with a thin WF inside it. WFs can have a holographic relief form. Then it will form an image in the reflected light, just like a regular hologram with a surface relief. However, unlike the latter, the part of the light beam that has passed through the wavelike hologram will not have a significant phase distortion.

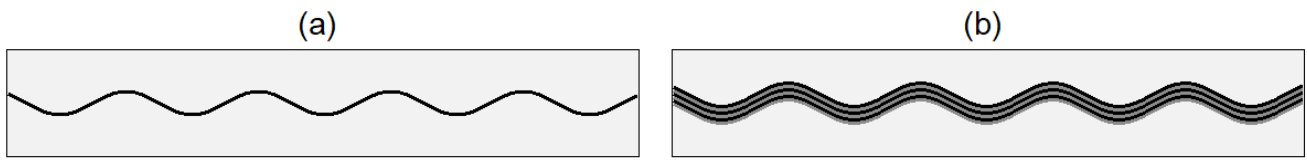

Fig. 1. OE with a semitransparent holographic (local periodic) WF embedded into the bulk of the OE: (a) single layer, (b) multilayer.

The OE with holographic WF can be used as a beam combiner when the image is projected onto the windshield (or onto a special screen between the windshield and the driver) of a vehicle. Windshield projection systems, or HUDs were invented quite a long time ago, first for use in aviation applications and then for automobiles [10,11,12]. However, only a small percentage of cars are equipped with such devices. Due to the widespread availability of GPS navigators in cars, which increase the risk of accidents by requiring the driver to move his gaze off the road and to the navigator screen changing the focal point of the eyes from far to near, the development of inexpensive screens for projection systems remains relevant. A projection system on the windshield would allow one to avoid such risks because the information is shown in front of the driver's eyes overlaid over the road in the background. The image plane is thus located either far or at infinity, i.e. for its observation shifting the focal point of the eye is not required.

Currently, volume holograms or surface phase relief holograms are used in HUDs. The disadvantage of volume holograms is that they cannot be replicated by embossing, which means they have a high cost. In addition, volume holograms recorded on traditional silver halide photographic materials are short-lived and change their characteristics depending on air humidity. Another disadvantage of conventional holograms, which is especially strong in the case of holograms with surface phase relief, are phase distortions of the wave passing through the hologram. Holograms containing single-layer metal WFs (Fig. 1a). are devoid of these drawbacks. These holograms can be used in a HUD for monochrome image projection. The color HUD will be discussed in the next section. Color projection will be very different when using waveguide displays. This option is discussed in the last section of the article.

Similarly an OE with holographic WF can be used in projection systems mounted on glasses (also known as HMDs, head-worn displays, wearable glasses, augmented reality glasses, smart glasses) [13,14,15,16,17]. HMDs were first developed for military aircraft (a projection system in a pilot's helmet). In this case, the projected image also appears at infinity. Through these glasses a user sees the real world. So, the transmitted wave distortion should be negligible. These systems will help solve the next pressing problem. Nowadays, various mobile devices are widespread: smartphones, tablets, laptop computers, etc. These devices often have a high quality display and are able to show a large amount of information. As the amount of information displayed increases, a larger display becomes desirable, which conflicts with the goal of preserving the compactness of the portable device. Smart glasses can solve this problem. We assume that the world is now on the verge of a boom of such systems.

For investigation of the OE feasibility let us estimate the maximum phase modulation of transmitted light. We will use the expression (38) from [1] for phase modulation $\delta$ of transmitted wave for a beam normally incident onto the OE containing a single-layer WF with a sinusoidal shape:

$$\delta = \frac{4\pi^3 h^2 d_1 n_0 (n_1 - n_0)}{\Lambda^2 \lambda n_1}, \qquad (1)$$

where $d_1$ is the WF thickness, $h$ and $\Lambda$ are the amplitude and period of the WF sinusoidal modulation correspondingly, $n_1$ is the refractive index of the WF, $n_0$ is the refractive index of transparent material of the OE in which the WF is embedded, $\lambda$ is the wavelength of the light.

For the creation of OEs with embedded WFs, some of the most useful materials are metal single-layer films with a thickness in the order of the inverse of the absorption coefficient of the metal (from a few nanometers to tens of nanometers). We estimated the maximum phase modulation for the typical parameters of these samples: $h$=0.2 μm, $d_1$=0.021 μm, $n_0$=1.5, $n_1$=0.97, $\lambda$=0.5 μm, $\Lambda$=2 μm. Here we use the refractive index of Au from [18] for $n_1$. Substitution of these parameters into (1) gives $\delta = 0.012\pi$. The estimated value of the phase modulation is sufficiently small, which allows one to consider the transmitted light as being practically undisturbed.

### 3. OE with holographic multilayer WF

Another type of OE shown in Fig 1b implies the presence of several parallel wavelike layers of materials with different refractive indices. Thin metallic films separated by dielectric layers can be used for the multilayer WF. This is similar to the model used in [19] for X-ray diffraction on a multilayer structure modulated by surface acoustic-waves. The same transparent material into which the multilayer WF is embedded can be used for the dielectric layers between metalic WF. This option enables obtaining a low value for the phase modulation of light transmitted through the OE with multilayer WF. A drawback of this option is that the diffraction efficiency of such WF will be lower than the diffraction efficiency of a fully dielectric WF described below due to light absorption in the metallic films. As another option, different types of dielectrics could be used in adjacent layers of the multilayer WF. Particularly, two alternate dielectric layers with thicknesses $d_1$ and $d_2$ can be used [20]. The period $(d_1+d_2)$ of multilayer film in direction of the optical axis can be estimated from the Bragg condition $2(d_1+d_2)\sin\theta_B=\lambda/n_{av}$, where $n_{av}$ is an average refractive index of multilayer film, $\theta_B$ is a Bragg angle, (see equation (9.10) in [21]). For $\theta_B=\pi/2$ $(d_1+d_2)=\lambda/2n_{av}$. So, the period of a multilayer WF significantly exceeds the thickness of the metal single-layer film. The entire thickness $N(d_1+d_2)$, of a multilayer WF, where $N$ is a number of pairs of alternate dielectrics layers, will be one or two orders of magnitude higher than the thickness of a single-layer WF. More precisely, the thickness of the multilayer WF will be obtained below.

Multilayer WFs have spectral selectivity and can be used to create a new type of high-quality color holograms representing a combination of three multilayer holographic WFs recorded with light beams of three different wavelengths (red, green, blue) and located inside a single layer of transparent material (see Fig. 2). The period in the direction of the optical axis for each of the holographic WFs is chosen so that each WF selectively reflects just the wavelength that was used during its recording. During reconstruction with white light, each of the three holographic WF cuts a portion of the spectrum within the band of its spectral selectivity. Reconstructing radiation, the wavelength of which is outside of the spectral selectivity of some multilayer WF, passes through it without significant phase distortions. The wave reconstructed by some WF passes through the other WFs also without significant phase distortions. Thus, a quality color image is reconstructed.

The proposed color holograms have several advantages over known types of color holograms. Denisyuk holograms [22] allow one to obtain very high quality color images. For full color rendition in the same recording medium with three different wavelengths, a superposition of three holograms is recorded. However, registering of the superposition of holograms in one photo layer leads to a decrease in the diffraction efficiency (section 17.6.3 in [21]). Therefore, to reconstruct such holograms, special particularly bright light sources are required. In addition, volume holograms are obtained only as a result of direct holographic registration, i.e. they do not allow mass mechanical replication. This leads to an increase in their cost. Other types of holograms that allow reconstruction by white light are rainbow holograms or Benton's holograms [23]. In these holograms, images with different wavelengths are visible at different angles in the vertical plane. Therefore, the correct color of the reconstructed image can be seen only with one strictly defined position of an observer. Such holograms are extremely widespread, due to their suitability for mechanical replication and, consequently, low cost. However, the image quality of holograms of this type is significantly lower than that of Denisyuk holograms. This is due to the use of additional optical elements (lenses and slits) as well as two-step registration. In addition, since the color of an image in holograms of this type changes with a change in the position of the eye of an observer, as a rule, the observer sees the image in relative colors. There is a reason to believe that the proposed color hologram will combine high image quality together with an ability to be replicated mechanically in large quantities, combining the advantages of Denisyuk and Benton holograms.

The OE with multilayer WF can be fabricated similarly to the fabrication process that was described in the Introduction. The difference is that instead of metal deposition, a sequence of dielectrics with different refraction indexes should be used for the deposition. This is similar to the technology of multilayer mirror fabrication. For color holograms all operations should be repeated three times. Another option is to fabricate three wavelike holograms (red, green, and blue) separately and then adhere them together. Both options for color hologram fabrication are similar to the corresponding technologies used for the manufacture of multilayer optical data discs [4].

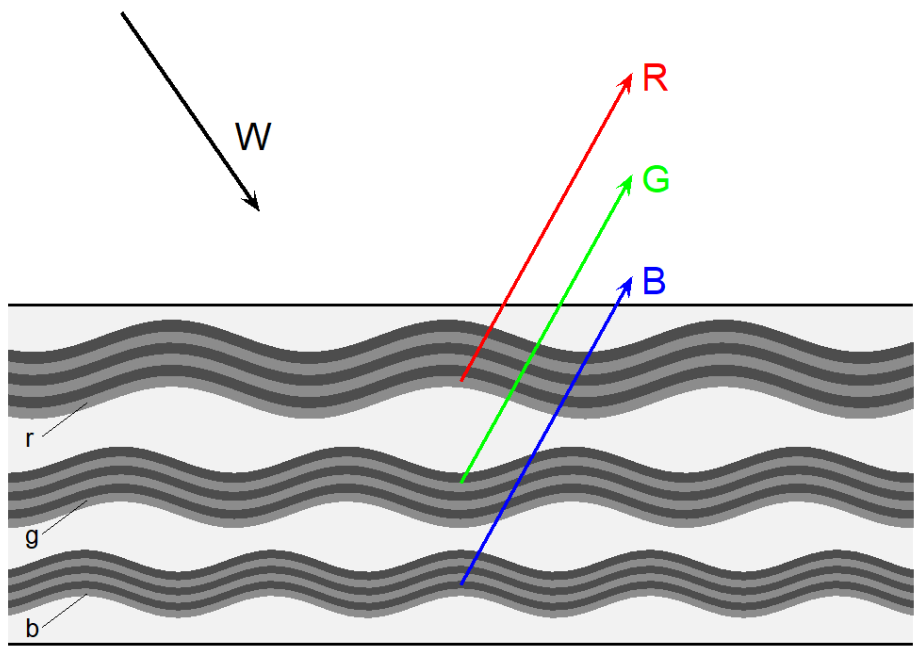

Fig. 2. Scheme of the proposed color hologram.
Three multilayer wavelike holograms r, g, and b recorded by different wavelengths (red, green, and blue, respectively) are embedded into a single transparent medium.
W is the reconstructing beam of white light; R, G, and B are reconstructed red, green, and blue light beams, respectively.

The parameters of the WF can be found with a computer simulation, which may show how many periods are necessary to obtain the desired values for the diffraction efficiency and spectral selectivity. Then, the phase modulation of light transmitted through the WF should be estimated. In computer simulations, we studied a structure containing three multilayer holographic WFs, corresponding to Bragg wavelengths of 442 nm, 529 nm and 647 nm and consisting of alternating layers with refractive indices $n_1$ and $n_2$. We calculated the diffraction efficiency of each multilayer WF as a function of a reconstructing wavelength (spectral selectivity). The diffraction efficiency is defined as [24]:

$$DE = \left|\frac{c_1}{c_R}\right| S_1 S_1^*, \qquad (2)$$

where $S_1$ is the amplitude of the 1-st order of diffraction by the wavelike structure, $c_R$ and $c_1$ are cosines of the reconstructing beam and the 1-st diffraction order to the direction of the optical axis, the asterisk means complex conjugation. We used the expression from [1] for the case of pure dielectric multilayer structure:

$$S_1 = -i\chi J_1(Ph)\left\{i\Gamma_1 + \frac{c_1\left[\gamma_1 \exp(\gamma_2 N(d_1+d_2)) - \gamma_2 \exp(\gamma_1 N(d_1+d_2))\right]}{\exp(\gamma_2 N(d_1+d_2)) - \exp(\gamma_1 N(d_1+d_2))}\right\}^{-1}, \qquad (3)$$

where

$$\gamma_1 = -i\frac{\Gamma_1}{2c_1} + \frac{1}{2}\left[-\left(\frac{\Gamma_1}{c_1}\right)^2 - \frac{4\chi^2 J_1^2(Ph)}{c_R c_1}\right]^{\frac{1}{2}}, \gamma_2 = -i\frac{\Gamma_1}{2c_1} - \frac{1}{2}\left[-\left(\frac{\Gamma_1}{c_1}\right)^2 - \frac{4\chi^2 J_1^2(Ph)}{c_R c_1}\right]^{\frac{1}{2}}, \qquad (4)$$

$\chi = \frac{\pi\varepsilon_1}{2\lambda\varepsilon_0^{1/2}}$, $\Gamma_1$ is the Bragg dephasing [1,24], $J_1(a)$ is the 1-st order Bessel function, $\varepsilon_0$ and $\varepsilon_1$ are the mean value and modulation of the dielectric constant, respectively, and $P = 2\pi/(d_1+d_2)$. Expressions (3, 4) are special cases of the general expressions which were obtained in [1], in the framework of the dynamic diffraction theory of multilayer WF, for the case when only one diffraction order is found in the angular selectivity band of the structure. The calculated spectral selectivity for the different numbers of the pairs of alternating layers $N$ and $n_1 = 1.65$, $n_2 = 1.38$ is shown in Fig. 3. The remaining parameters were selected from the condition that the angles of inclination of the incident and diffracted beams for all three WFs are the same and have cosines $c_R = -0.81$ and $c_1 = 0.69$. These angles determine the wave number $K = 2\pi/\Lambda$ of each hologram related to the sinusoidal modulation of the WF. For holograms corresponding to wavelengths of 442, 529, and 647 nm, $K$ is $2.9\times10^6$, $2.4\times10^6$, and $2.0\times10^6$ m$^{-1}$, respectively. The periodicity of multilayer WF along the optical axis was determined from the Bragg condition. We assumed that $d_1 = d_2$ and are equal to $9.7\times10^{-8}$, $1.2\times10^{-7}$, and $1.4\times10^{-7}$ m for the holographic WFs corresponding to wavelengths of 442, 529, and 647 nm, respectively. The amplitude $h$ of the sinusoidal modulation along the WF was chosen to be $5.2\times10^{-8}$, $6.3\times10^{-8}$ m, and $7.5\times10^{-8}$ m for holographic WFs corresponding to wavelengths of 442 nm, 529 nm, and 647 nm. The refractive index of the medium in which three holographic WFs were placed was equal to the average value of the refractive indices of the alternating layers: $n_0 = 1.52$.

To estimate the maximum modulation of the phase of light of wavelength $\lambda$ that is passing through the considered OE, we will use the expression obtained in [1]:

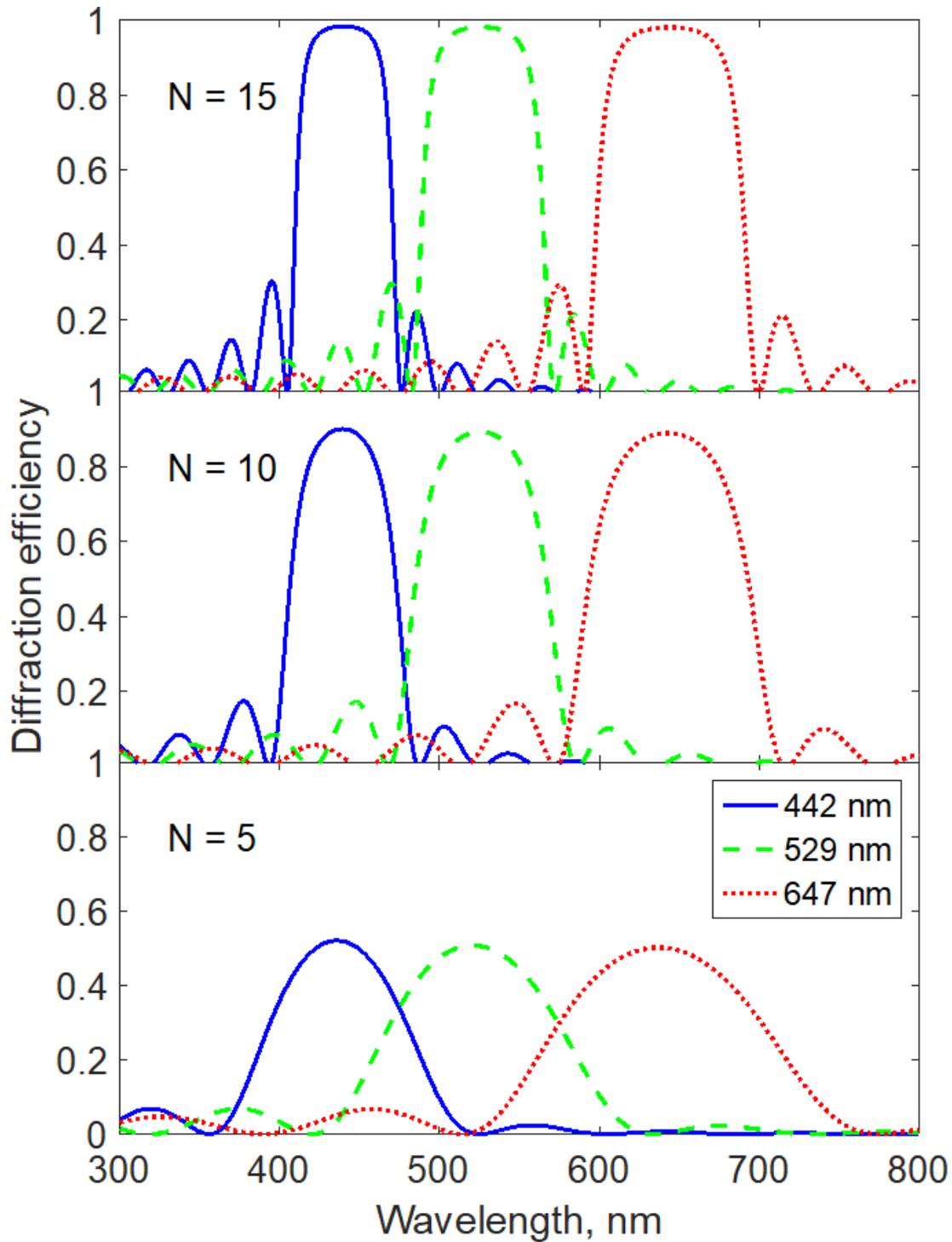

Fig. 3. The calculated spectral selectivity of a color hologram containing three multilayer holographic WFs corresponding to Bragg wavelengths of 442 nm, 529 nm, and 647 nm for the different numbers of the pairs of alternating layers N. The parameter values of the WFs are given in the text.

$$\delta = \frac{4\pi^3 h^2 n_0 N}{\Lambda^2 \lambda} \left[ \frac{d_1(n_1 - n_0)}{n_1} + \frac{d_2(n_2 - n_0)}{n_2} \right], \tag{5}$$

We should note that the terms in square brackets, corresponding to the phase shifts related to the adjacent layers, have different signs, and partially compensate each other. This permits a low value of δ for a relatively large thickness of the entire WF. Substitution of the structure parameters described above results in a value of δ equal to -0.0031π for each of the three holographic WFs.

Such a value for the maximum phase modulation allows us to consider the level of distortions of the transmitted wave as insignificant. This demonstrates the principal possibility of realizing the proposed type of color holograms. Such holograms can also be used as a screen in color projection systems HUD and HMD mentioned in section 2.

## 4. OE with diffuse WF

A single-layer WF may also have a diffuse chaotic relief shape (see Fig. 4a). The OE with WF that have random chaotic reliefs diffusely scatter reflected light while the transmitted light phase is not disturbed. The relief of such WFs may contain both micro asperities and larger elements that can be distinguished with the naked eye. The relief can be obtained by taking a replica of a diffusely reflecting surface, for example, engraving metal, fabric surface, etc. Thus, the obtained OE will exactly simulate for an observer the texture of the original surface. Additionally, a unique visual effect will take place because the OE with diffuse WF is semi-transparent, while the origin surface is opaque. The multilayer diffuse WF (Fig. 4b) is also possible. It will have the spectral selectivity similarly to the multilayer holographic WF.

Possible applications of OEs with diffuse WF are described next. Sometimes it is desirable to close off a certain area from external observations, while retaining the ability to observe from this area what is happening outside. This application is based on the assumption that the intensity of the luminous flux falling on the WF outside is significantly higher than from the enclosed area. This can be done with a help of special OEs containing diffuse WFs, which are called one-way shields or privacy windows (Fig. 5). If the difference in the intensity of the luminous flux is insufficient, then an additional backlight can be used outside. In this case, a powerful collimated beam of light should be directed so as not to get into the eye of the observer from the inside. For example, it can be directed to the floor or to the ceiling where light absorbing screens are installed.

Similarly, diffuse WFs can be used as an alternative to tinting car windows. In this case, choosing the thickness of the metal WF one can get any desired transmittance coefficient. On the other hand, the choice of a suitable diffuse image on the glass of a car helps to hide those in the cabin from an outside observer. This can be achieved with a higher transmittance than conventional tinted glass. Thus, the proposed technology can improve traffic safety. In addition, it opens up new opportunities for car design. Similarly, diffuse WFs can be used instead of mirrored glass in high comfort buildings. In this case, diffuse images visible in the reflected light can both play the role of building design elements and also can be used for outdoor advertising. Additionally, OEs containing diffuse WFs can be used as a semitransparent screen for projecting advertising images onto it. These images will not be seen from the back side of the screen. This would allow its use in shop-windows of boutiques in big shopping centers.

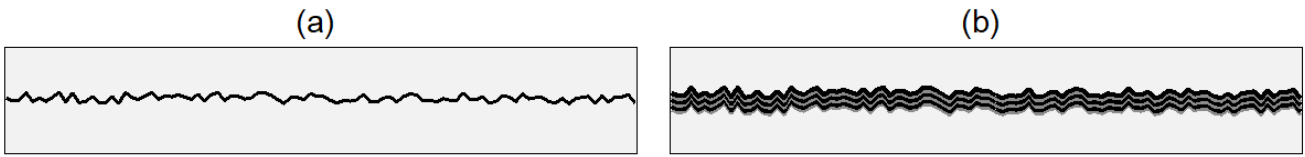

Fig. 4. OE with a semitransparent diffuse chaotic WF embedded into the bulk of the OE: (a) single layer, (b) multilayer.

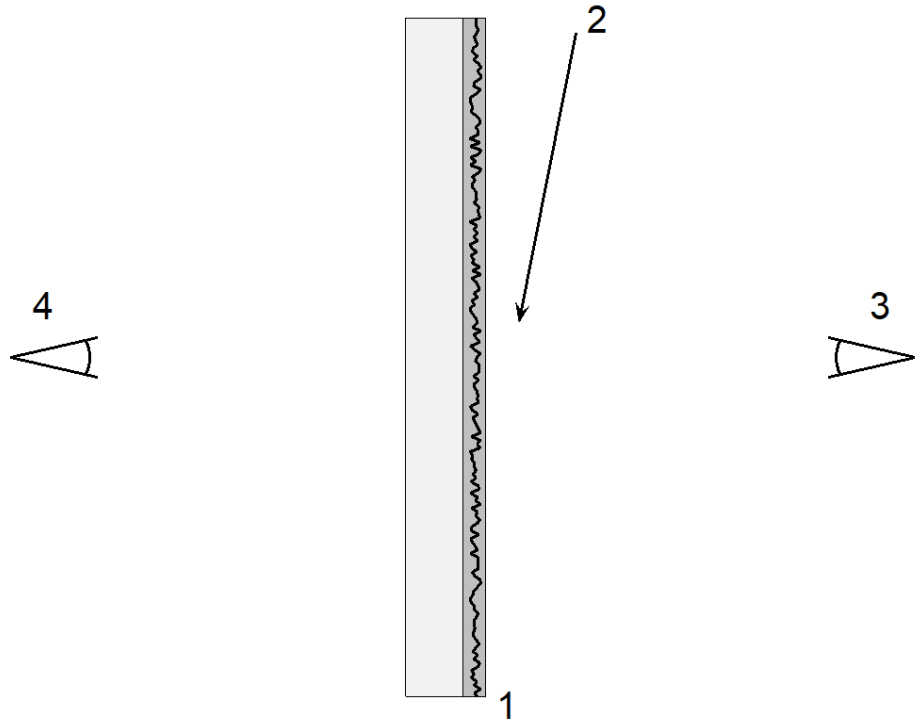

Fig. 5. Privacy window. 1 – OE with diffuse WF, 2 – back-light, 3 – outside observer, 4 – inside observer.

Note here the well-known technology of applying one-way images to a perforated film made of an opaque material, which also allows placing images visible only from the outside on the windows of vehicles (see, for example, patents [25,26]). However, this technology is more suitable for outdoor advertising than for car window decoration.

Another possible use of diffuse WFs could be images on eyeglasses. In particular, they can be a trademark of the manufacturer. Such an image will be visible to an external observer, but will remain invisible to the wearer of the glasses. In another case, such image can serve as an element of design of the glasses. The method can be used both for vision correction glasses and sunglasses.

Using OEs with multilayer diffuse WFs enables adding color to all the applications mentioned in this section. Additionally, these WFs can model the fabric surface texture and simultaneously have spectral selectivity. This can be used for the following special effects. These OE will be visible only when their illumination lies within WF's spectral selectivity band. Otherwise, the WF is invisible. This can be used in theatrical shows for details on curtain or actor's clothing to appear or disappear when subjected to illumination.

The OEs with diffuse single layer WFs can be fabricated similarly to the fabrication of the OE with holographic single-layer WF that was described in the previous chapter. The only difference is that instead a master hologram recording we should take a replica of some random relief surface. Then this replica is used for fabrication of the master stamp. The diffuse multilayer WF can be fabricated similarly to the fabrication of the holographic multilayer WF.

Now, let us estimate the maximum phase modulation of light transmitted through the OE with diffuse WF. From expression (39) derived in [1] for a beam normally incident onto the OE containing a multilayer WF with arbitrary shape it is possible to obtain the following expression:

$$\delta = \frac{\pi n_0 N(\psi)^2_{max}}{\lambda} \left[ \frac{d_1(n_1 - n_0)}{n_1} + \frac{d_2(n_2 - n_0)}{n_2} \right], \tag{6}$$

where $\psi$ is the angle between the local normal to the surface of WF and the optical axis. A specific value of $\delta$ will be obtained from (6), in the next section, using the experimental data.

## 5. Experimental

Experimental specimens of OEs containing aluminum WFs were fabricated in a private laboratory of O.B. Serov using UV curing polymers. The thickness of the aluminum film was about 2 nm. This thickness is on the order of the inverse of the absorption coefficient of the aluminum, which provides a high enough intensity for both transmitted and reflected light. The OE fabrication method was as follows: firstly, we produced the polymer replica of an object surface. In one instance, we used the surface of ordinary holograms with a phase relief as the object surface. In another instance, we used a chaotic diffuse surface. Then, we sputtered the Al onto the polymer replica. Then, we placed a slab of UV cured polymer on the Al layer and covered it by a piece of glass. The OE with a chaotic diffuse WF perfectly imitated the source surface texture for an observer. We estimated the maximum phase modulation holographic wavelike film for the typical parameters of these samples: $h = 0.2$ μm, $d_1 = 0.002$ μm, $n_0 = 1.5$, $n_1 = 0.81$, $\lambda = 0.5$ μm, $\Lambda = 2$ μm. The refractive index $n_1$ of Al was taken from [27]. The substitution of these parameters into expression (1) gives $\delta = 0.007\pi$. This is a very small amount. In contrast, in ordinary reflection holograms with phase relief the optical path length of the rays passing through the hologram are varied up to $\Delta h(n_0 - 1)$, where $\Delta h$ is the change of depth of the surface relief, and $n_0$ is the refraction index of the hologram material. For this case typically $\Delta h = \lambda/4$ and $\delta \sim 0.25\pi$. This amount of the phase modulation causes a significant distortion of the beam passing through the hologram.

The OE with a chaotic diffuse WF was experimentally investigated as follows. We studied a specimen containing several parts with the upper polymer slab above a chaotic diffuse wavelike film and several parts without the upper slab. The intensity of scattered radiation transmitted through the OE was measured. We directed the laser beam with 0.6 μm wavelength normally onto the specimen. In Fig. 6a the angular diagram of scattered light behind the sample is shown. The intensity of scattered light in a certain direction divided by the incident beam intensity is shown as a function of the angle between this direction and the optical axis. Curve A relates to the parts of the specimen with the upper polymer slab and curve B relates to the parts of the specimen without the upper slab. The intensity of the directly transmitted beam (the angle is equal 0) for curve A is equal to 0.83. This is almost 30-fold higher than the intensity of the directly transmitted beam for curve B (0.029). The intensity of the scattered beam for curve A is significantly lower than the one for curve B. This measured data demonstrates the amount of reduction of transmitted light scattering for the chaotic diffuse wavelike film fully embedded into polymer layer.

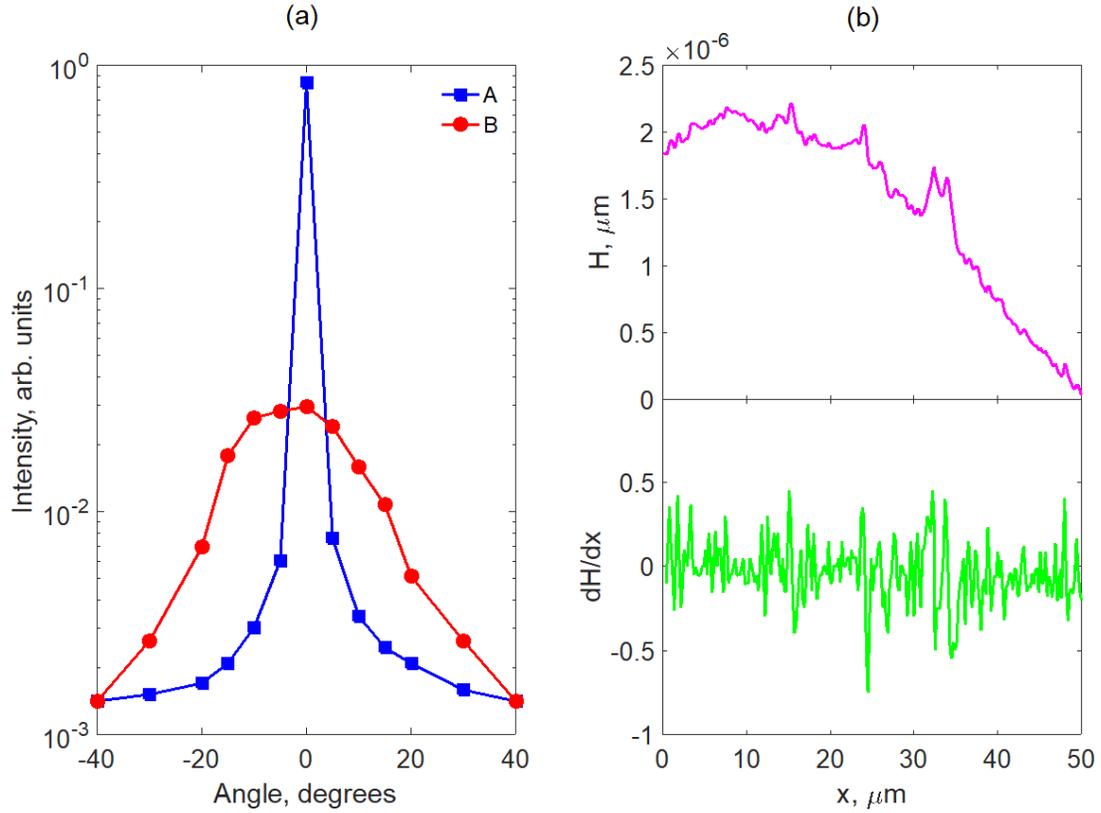

Fig. 6. Experimental specimen containing diffuse aluminum WF. (a) The intensity of scattered light in a certain direction as a function of the angle between this direction and the optical axis. Curve A relates to the parts of the specimen with the upper polymer slab and curve B relates to the parts of the specimen without the upper slab. (b) Relief cut (upper graph) and its first derivative (lower graph) measured by an atomic force microscope.

For estimation of the phase modulation of the experimental specimen containing diffuse Al WF using expression (6), we should know the value of $(\psi)^2_{max}$. For this, we also used an experimental sample with chaotic diffuse WF containing some areas without the upper transparent polymer layer. Several areas of 50×50 μm$^2$ were scanned with an atom force microscope. One of the representative cross-sections of these scans is shown on the upper graph of Fig. 6b. The value of $\psi$ corresponding to the local relief incline is not clearly seen from this graph as the scales of vertical and horizontal axes are very different. However, $\psi$ can be found from the expression: $\tan(\psi) = dH/dx$, where $H$ is the local relief height, $x$ is its horizontal coordinate (we consider a relief cut by the plane containing the optical axis). The lower graph of Fig. 6b shows the first derivative of the relief shape. This graph allows us to use for $(\psi)^2_{max}$ value equal to 0.4. We used the following values of the other parameters: $d_1$ = 0.007 μm, $n_0$ = 1.5, $n_1$ = 1.26, $\lambda$ = 0.6 μm. Here we use the refractive index of Al from [27] for $n_1$. For a single-layer WF, using N = 1 and substituting these parameters into the expression (6), excluding the second term in square brackets, results in $\delta = 0.0015\pi$. Such a low value of $\delta$ confirms the possibility of producing the proposed type of OEs.

## 6. Discussion

Let us discuss the prospects for using the proposed types of OEs containing WFs. We are convinced that their application should be started with OEs containing metal single-layer WFs. First, this is due to the fact that the technology of their production practically does not differ from the well-developed technology for the production of replicated holograms and can be produced on already existing equipment. Note that the cost of such production is very low. At the same time, it is now difficult to estimate what the production cost of OEs containing multilayer WFs will be. In addition, the constraints obtained in this work on the parameters of multilayer wavy films that do not distort the transmitted wave are rather rigid. At the same time, OEs containing metal single-layer wavy films, with any reasonable parameters, will not distort the phase of the transmitted wave.

OEs containing metal single-layer WFs are of greatest interest for HUD and HMD as so-called beam combiners. At first glance, it seems that they can only be used there in monochrome projection. Indeed, in a conventional volumetric projection of color (RGB) image, you must use three beam combiners (one for each wavelength) with different grating periods. In particular, for this use-case, one can use the color hologram proposed in this paper, containing three spectrally selective multilayer WFs (Fig. 2).

However, in waveguide displays [11,12,14,15,16,17], for output a color image from a waveguide, one grating is sufficient for all three wavelengths. A schematic diagram of a waveguide projection suitable for both HUD and HMD is shown in Fig. 7 (this scheme is similar to that used in [16], but differs from it in the use of reflective holograms instead of transmission ones). The scheme includes a planar waveguide WG, a hologram $H_1$ for input the light beam PR from the projector into the waveguide, and a hologram $H_2$ for output light from the waveguide towards the observer OB. In addition, the observer sees through the waveguide and the $H_2$ hologram the image of the "real world" RW. When using this scheme in an automobile HUD, it is convenient to place an $H_2$ hologram inside a film located in the middle of a laminated triplex windshield. The magnification of the projected image is achieved due to the fact that the size of the $H_2$ hologram in the direction of propagation of the light flux in the waveguide (in Fig. 7 from left to right) significantly exceeds the size of the $H_1$ hologram. Achromatization in this scheme is obtained due to the fact that the holograms $H_1$ and $H_2$ have the same spatial frequency. Inside the waveguide, the rays propagate in a zigzag manner, due to total internal reflection at the boundaries. In this case, beams of different wavelengths have different slopes (this is not shown in Fig. 7).

Usually, volume gratings are used as holograms $H_1$ and $H_2$. Holograms of this type are poorly suited for mass production. They cannot be replicated by embossing, but need to be recorded in a holographic experiment. As a result, volume holograms have a high cost. In contrast, holograms containing single-layer metal WFs (Fig. 1a) can be replicated using traditional embossing technology at a low cost.

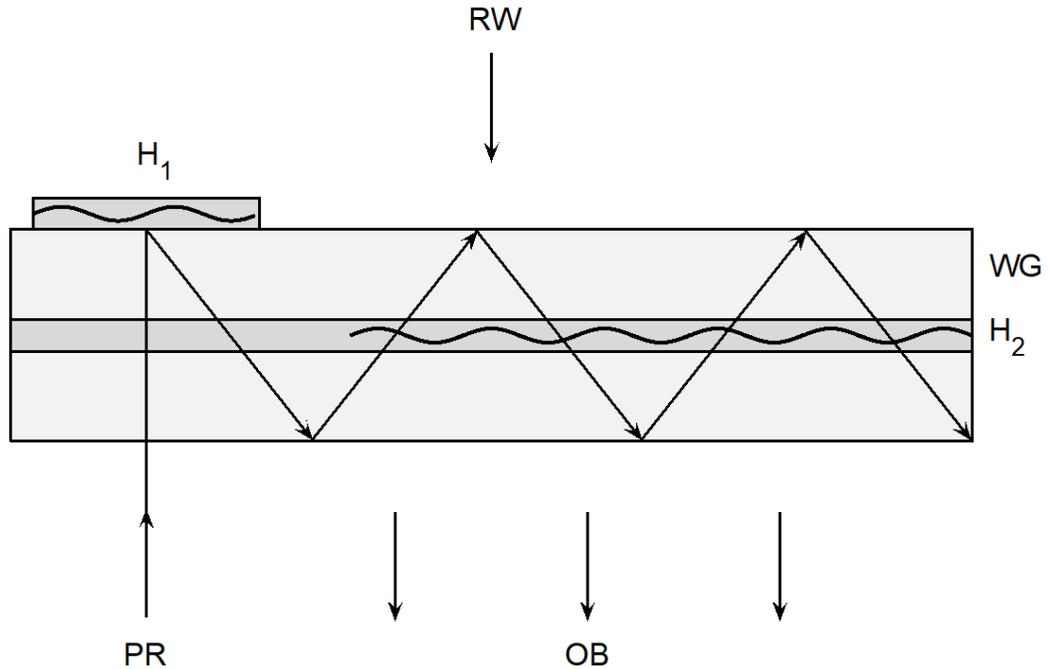

Fig. 7. Waveguide display. WG – waveguide, H1 – input hologram, H2 – output hologram, PR – beam from projector, RW – real world, OB – observer.

There is one more problem. The $H_2$ hologram should not distort the "real world" seen by the user through it. One of the advantages of the waveguide version of the display is that the rays incident on the $H_2$ hologram at angles not too far from the normal are diffracted into the waveguide mode and do not enter the user's eyes. However, the result of the diffraction of rays falling at large angles can produce stray light that hits the eyes. For both HUD and HMD, such oblique and sufficiently bright rays can come from the sun or artificial light sources. Here, the use of OEs containing metal single-layer WFs will avoid the described unpleasant effect, since it was shown above that they do not introduce phase distortions into the transmitted wave. The estimates obtained in the second section are perfectly suitable for this case.

In addition, there is the following problem. When the radiation of the waveguide mode is diffracted by the hologram $H_2$, the diffracted part of radiation leaves the waveguide, and the part of radiation that has passed through the hologram without diffraction (zero order) remains in the waveguide. This radiation will gradually move along the waveguide, reflecting from its boundaries, and will be output by the hologram from the waveguide as a result of diffraction. In this case, the waveguide beam will gradually weaken. To ensure uniform brightness of the image over the area of the hologram, its diffraction efficiency should not be uniform over the area of the hologram, but should increase in the direction of the displacement of the waveguide mode radiation [17]. Obtaining a hologram with diffraction efficiency inhomogeneous over the area causes certain technological difficulties, and further increases the cost of the hologram. When using holograms containing single layer metal WFs, there are two ways to solve the problem. In the first option, the grating relief remains uniform over the area of the hologram, and the thickness of the metal layer sprayed onto the relief gradually increases from one edge of the hologram to the other. In the second option, the original hologram is recorded with an inhomogeneous relief height. In this case, the complexity of manufacturing for only the so-called master holograms grows. But the increase of the cost of each hologram-copy during mass replication using the embossing method is insignificant.

## 7. Conclusion

OEs containing holographic, diffuse, single-layer, and multi-layer WFs were analysed for their possible applications. The phase distortion of a wave transmitted through the different types of OE containing WF was estimated. The spectral selectivity of the OE containing three multilayer WFs corresponding to the different Bragg wavelengths was calculated. The microrelief of the experimental specimen containing a diffuse Al WF was measured by atomic force microscope. The result of the measurement was used for an estimation of the transmitted light phase modulation. The angular diagram of scattering light transmitted through the experimental specimen was measured. The scheme of a waveguide display with a single-layer holographic WF was analyzed.

## Acknowledgments

The work was carried out within the framework of the state task (Russia) and in the framework of the Memorandum of understanding between Kotel'nikov Institute of Radio Engineering and Electronics of the RAS, Moscow, Russia and University of Sonora, Hermosillo, Mexico. We thank A.P. Orlov (IRE RAS, Moscow, Russia) for his help in editing the graphs in Fig. 6b, G. Chernov (Universidad de Sonora, Hermosillo, Sonora, México) for his help with language corrections.

## References

1. Smolovich A.M., Chernov V. Optical elements containing semitransparent wavelike films. *Applied Optics*. 2017. V. 56. № 22. P. 6146–6155. Erratum: *Applied Optics*. 2018. V. 57. № 30. P. 8914-8914. https://doi.org/10.1364/AO.56.006146
2. Patent USA No 4315665. Haines K.A. *Composite optical element having controllable light transmission and reflection characteristics*. 1982.
3. Patent USA No 9244201. Dillon S.M. *Diffuse reflecting optical construction*. 2016.
4. Ichimura I., Saito K., Yamasaki T., Osato K. Proposal for a multilayer read-only-memory optical disk structure. *Appl. Opt.* 2006. V. 45. № 8:1794–1803. https://doi.org/10.1364/AO.45.001794
5. Smolovich A.M., Cervantes M.A., Chernov V. Multilayer optical disk and method of its management for preventing its illegal use. Proceedings of SPIE - The International Society for Optical Engineering. 2007. https://doi.org/10.1117/12.738647
6. Smolovich A.M. Achromatic optical elements. *Appl. Opt.* 2006 V. 45. № 30. P. 7871–7877. https://doi.org/10.1364/AO.45.007871
7. Smolovich A.M. Geometrooptical mechanism of wave-front reconstruction. *Opt. Spectrosc*. 2020. V. 128. № 9. P. 1393–1400. https://doi.org/10.1134/S0030400X20090209
8. Zacharovas S., Bakanas R., Bulanovs A., Varadarajan V. Effective public security features for embossed holograms. *Proceedings of SPIE - The International Society for Optical Engineering*. 2017. P. 1012702-10. https://doi.org/10.1117/12.2248904
9. Johnston S. Holograms. *A Cultural History*. Oxford University Press. 2016. 271 p.

10. Peng H., Cheng D., Han J. et al. Design and fabrication of a holographic head-up display with asymmetric field of view. *App. Opt*. 2014 V. 53. № 29. P. H177–H185. https://doi.org/10.1364/AO.53.00H177
11. Draper C.T., Bigler C.M., Mann M.S., Sarma K., & Blanche P. A. Holographic waveguide head-up display with 2-D pupil expansion and longitudinal image magnification. *Appl Opt*. 2019. V. 58. № 5. P. A251– A257. https://doi.org/10.1364/AO.58.00A251
12. Gu L., Cheng D., Wang Q., Hou Q., Wang S., Yang T., & Wang Y. Design of a uniform-illumination two-dimensional waveguide head-up display with thin plate compensator. *Optics Express*. 2019. V. 27. № 9. P. 12692–12709. https://doi.org/10.1364/OE.27.012692
13. Cakmakci O., Rolland J. Head-worn displays: A review. *IEEE/OSA Journal of Display Technology*. 2006. V. 2. № 3. P. 199–216. https://doi.org/10.1109/JDT.2006.879846
14. Lee B., Lee S., Cho J., Jang C., Kim J., and Lee B. Analysis and implementation of hologram lenses for see-through head-mounted display. *IEEE Photonics Technology Letters*. 2017. V. 29. № 1. P. 82–85. https://doi.org/10.1109/LPT.2016.2628906
15. Zhan T., Yin K., Xiong J., He Z., & Wu S. T. Augmented reality and virtual reality displays: perspectives and challenges. *iScience*. 2020. V. 23. № 8. P. 101397. https://doi.org/10.1016/j.isci.2020.101397
16. Koreshev S.N., Shevtsov M.K. Holographic sight of the light-guide type with a synthesized pupil. *J. Opt. Technol*. 2018. V. 85. № 3. P. 153–156. https://doi.org/10.1364/JOT.85.000153
17. Putilin A.N., Morozov A.V., Kopenkin S.S., Dubynin S.E. & Borodin Yu.P. Holographic waveguide periscopes in augmented reality displays. *Opt. Spectrosc*. 2020. V. 128. № 11. P. 1828–1836. https://doi.org/10.1134/S0030400X2011020X
18. Johnson P.B., Christy R.W. Optical constants of the noble metals. *Physical Review B*. 1972. V. 6. № 12. P. 4370–4379. https://doi.org/10.1103/PhysRevB.6.4370
19. Erko A.I., Roshchupkin D.V., Snigirev A.A., Smolovich A.M., and Nikulin A.Yu. X-ray diffraction on a multilayer structure modulated by surface acoustic waves. *Nuclear Inst. and Methods in Physics Research, A*. 1989. V. 282. № 2–3. P. 634–637. https://doi.org/10.1016/0168-9002(89)90067-3
20. Sisakyan I.N., Smolovich A.M. Diffraction upon sinusoidally modulated multilayer structure. *Computer Optics*. 1989. №. 6. P. 6–9 (in Russian).
21. Collier RJ, Burckhardt CB, Lin LH. *Optical Holography*. Vol. 19712. New York. Academic Press; 1971. 605 p.
22. Denisyuk Y.N. On the reflection of optical properties of an object in a wave field of light scattered by it. *Soviet. Phys. Doklady*. 1962. V. 7. № 6. P. 543–545.
23. Benton S.A. Holographic displays-A review. *Optical Engineering*. 1975. V. 14. № 5. P. 402-407. https://doi.org/10.1117/12.7971805
24. Kogelnik H. Coupled wave theory for thick hologram gratings. *The Bell System Technical Journal*. 1969. V. 48. № 9. P. 2909–2947. https://doi.org/10.1002/j.1538-7305.1969.tb01198.x
25. Patent USA No 7624524. Mullins J.M. *Self-adhering perforated display assembly*. 2009.
26. Patent USA No 9469081. Hill G.R. *Open perforated material*. 2016.
27. Refractive index database, http://refractiveindex.info/. 2020 (accessed July 24, 2021).